\documentclass[runningheads]{llncs}
\usepackage{graphicx}
\usepackage[ruled,vlined]{algorithm2e}
\begin{document}
\title{NeXtQSM - A complete deep learning pipeline for data-consistent quantitative susceptibility mapping trained with hybrid data \thanks{The source code will be provided upon request.}}
\titlerunning{NeXtQSM}

\author{
Francesco Cognolato\inst{1,2} \and
Kieran O’Brien\inst{2,4} \and
Jin Jin\inst{2,4} \and
Simon Robinson\inst{1,5,6,7} \and
Frederik B. Laun\inst{8} \and
Markus Barth\inst{1,2,3} \and
Steffen Bollmann\inst{1,2,3}
}
\authorrunning{F. Cognolato et al.}

\institute{
Centre for Advanced Imaging, The University of Queensland, Brisbane, Australia \and
ARC Training Centre for Innovation in Biomedical Imaging Technology, The University of Queensland, Brisbane, Australia \and
School of Information Technology and Electrical Engineering, The University of Queensland, Brisbane, Australia \and
Siemens Healthcare Pty Ltd, Brisbane, Queensland, Australia \and
High Field MR Center, Department of Biomedical Imaging and Image-Guided Therapy, Medical University of Vienna, Vienna, Austria \and
Department of Neurology, Medical University of Graz, Graz, Austria \and
Karl Landsteiner Institute for Clinical Molecular MR in Musculoskeletal Imaging, Vienna, Austria \and
Institute of Radiology, University Hospital Erlangen, Friedrich-Alexander-Universität Erlangen-Nürnberg (FAU), Erlangen, Germany
}
\maketitle

\begin{abstract}
Deep learning based Quantitative Susceptibility Mapping (QSM) has shown great potential in recent years, obtaining similar results to established non-learning approaches. Many current deep learning approaches are not data consistent, require in vivo training data or solve the QSM problem in consecutive steps resulting in the propagation of errors. Here we aim to overcome these limitations and developed a framework to solve the QSM processing steps jointly. We developed a new hybrid training data generation method that enables the end-to-end training for solving background field correction and dipole inversion in a data-consistent fashion using a variational network that combines the QSM model term and a learned regularizer. We demonstrate that NeXtQSM overcomes the limitations of previous deep learning methods. NeXtQSM offers a new deep learning based pipeline for computing quantitative susceptibility maps that integrates each processing step into the training and provides results that are robust and fast.
\end{abstract}

\section{Introduction}
Quantitative Susceptibility Mapping (QSM) is a Magnetic Resonance Imaging (MRI) technique which has gained a lot of attention in the last decade because of its potential to
extract in vivo magnetic susceptibilities \cite{https://doi.org/10.1002/nbm.3569,FoundationsofMRI,shmuelikarin}. The quantitative information contained in each voxel linearly reflects the tissue magnetic susceptibility. As a result, QSM is able to reveal information about iron concentrations in the gray matter \cite{Acosta-Cabronero364,vanBergen789,ironconc,SCHWESER20112789,ironconc2,ironconc3,LAMBRECHT2020113314}, demyelinating lesions in the white matter \cite{https://doi.org/10.1002/mrm.25189,myelinqsm,LAMBRECHT2020113314}, copper accumulation \cite{copperqsm}, blood oxygenation \cite{oxygenationqsm}, microbleeds \cite{bleedsqsm} and differentiates them from microcalcifications \cite{https://doi.org/10.1118/1.3481505}. Alterations in tissue susceptibility can be linked to ageing \cite{Acosta-Cabronero364} as well as neurological diseases such as Parkinson's disease \cite{parkinsonqsm,10.1371/journal.pone.0162460}, Alzheimer's disease \cite{10.1371/journal.pone.0081093}, Huntington’s disease \cite{vanBergen789} and multiple sclerosis \cite{MS-QSM,https://doi.org/10.1002/mrm.25420,MS-QSM2}. \\

The extraction of tissue susceptibilities requires multiple steps in a processing pipeline starting from the phase signal of a gradient-recalled echo (GRE) sequence and includes phase unwrapping, background field removal and dipole inversion (for a review see: \cite{FoundationsofMRI}). The dipole inversion extracts the susceptibility values and requires the solution of an ill-posed inverse problem. One solution proposed to overcome this ill-posed problem is known as ``Calculation of susceptibility through multiple orientation sampling'' (COSMOS) \cite{COSMOS,COSMOS2} which requires at least three different orientations to eliminate the singularities in the dipole operation. Despite the high quality of the COSMOS reconstruction, this method is not practical because of its long acquisition time and requiring the measurement of the object in differing rotations. \\

Due to the impracticability of COSMOS, single orientation solutions have been proposed for phase images. Besides traditional computer vision optimization methods, the recent advent and huge improvements in deep learning techniques \cite{DBLP:journals/corr/RussakovskyDSKSMHKKBBF14} led to an application of neural networks to multiple problems in MRI and have shown effectiveness in solving QSM problems such as background field removal \cite{BOLLMANN2019139,10.1007/978-3-030-32248-9_10} and dipole inversion \cite{DBLP:journals/neuroimage/BollmannRKBOPOL19,YOON2018199,chen2019qsmgan,lai2020learned,Gao_2020} (for a review see: \cite{https://doi.org/10.1002/nbm.4292}). \\
However, a limitation of the current state-of-the-art techniques is that the background field correction and dipole inversion are either treated as two independent problems (e.g. \cite{feng_modl-qsm_2021} does not include background field correction) leading to accumulations of errors between consecutive steps, or do not include a dipole inversion that delivers data consistent solutions (e.g. \cite{autoqsm}). \\
Due to the fact that these steps are mostly treated independently, any imprecision or inaccuracies can propagate to the next step affecting the final reconstruction. Total field inversion \cite{totalfieldinversion} and Total Generalised Variation (TGV) QSM \cite{tgvqsm} solve consecutive steps in a single optimization procedure to reduce error propagation. These methods require substantial amounts of compute time due to the difficult optimization problems limiting their application in vivo.\\

To overcome the computational cost, deep learning methods have been proposed that combine consecutive steps such as the dipole inversion and background field correction \cite{autoqsm,liu2019deep}, but these models do not incorporate data-consistency constraints in the dipole inversion. Achieving a data consistent solution would make the reconstruction more robust and invariant to the dipole kernel magnetic field direction and has been shown to be a beneficial post-processing step for deep learning solutions \cite{zhang2019fidelity}. Lastly, most current deep learning QSM techniques require in vivo training data. Often they utilise COSMOS \cite{COSMOS} data as a target, which is obtained by multi orientation measurements. Besides the lack of practicality in obtaining such large data sets for training, this is not a realistic target for a single orientation measurement. \\
To address these limitations, we propose a variational NEural network trained on compleX realistic strucTures for Quantitative Susceptibility Mapping (NeXtQSM). NeXtQSM is an end-to-end deep learning pipeline trained on hybrid data to solve the background field correction and dipole inversion in a data consistent fashion composed by two models which are trained together to solve the two processing steps jointly. Our study is evaluated on data from the QSM reconstruction challenge 2.0 \cite{https://doi.org/10.1002/mrm.28716} and on in vivo 7T data.

\section{Theory}
\subsection{QSM forward operation}
The QSM physical operation is generally described by the equation $Y = \Phi X + \epsilon$, where $X$ is the object in image space containing the susceptibility values, $\epsilon$ the noise during the measurement, $Y$ the measured local field and the forward operation $\Phi = F^{-1} D F$ which is a multiplication with the dipole kernel in Fourier domain. \\
For numerical efficiency, the convolution operation in image space is commonly calculated in Fourier domain $F$ as a pointwise multiplication. The dipole kernel in Fourier space for magnetic fields \cite{https://doi.org/10.1002/cmr.b.10083} in the z-axis can be represented with the following equation:

\begin{equation}
    D = \frac{1}{3} - \frac{k_{z}^{2}}{k_{x}^{2} + k_{y}^{2} + k_{z}^{2}}
    \label{dipolekernel}
\end{equation}

where $k_{x}$, $k_{y}$ and $k_{z}$ are the k-space values in the respective directions. When the k-space values fraction approaches 1/3, the dipole kernel results to 0, making the dipole inversion ill-posed.

\subsection{Variational Networks}
These sophisticated models are the bridge connecting iterative methods and deep learning \cite{10.1007/978-3-319-66709-6_23}. In iterative methods, the solution of inverse problems can be formulated as an optimization process where at each iteration an objective function $E(x)$ composed by two different terms is minimized:

\begin{equation}
    \arg \min_{x} E = \lambda \underbrace{||Y - \Phi x||_2^2}_\textrm{$f(x)$} + \Psi(x)
\end{equation}

where $f(x)$ is the data consistency term. In this specific case containing the QSM forward model $\Phi$, which helps in applying the model to a variety of input data that would lead to unstable solutions. $\Psi(x)$ is the regularizer or prior knowledge which prevents overfitting on the data and $\lambda$ is the learnable trade-off factor between data term and regularizer. \\
In iterative methods, a hand-crafted prior term $\Psi$ such as Total Variation \cite{totvar} is chosen. Whereas in variational networks, which can be considered as a hybrid iterative method, the regularizer is learned using convolutional neural networks. \\
At the end of the prefixed number of iterations, a reconstruction error $E_{recon}$ is calculated across the samples in the batch for updating the weights of the neural network:

\begin{equation}
    E_{recon} = \sum_{n=0}^{N} \frac{1}{N} ||\hat{x}_{S} - X||_2^{2};
\end{equation}

where $\hat{x}_{S}$ is the reconstructed susceptibility map and X is the actual ground truth susceptibility values obtained from the hybrid structures.

\begin{figure*}[!h]
    \centering
    \includegraphics[width=1\textwidth]{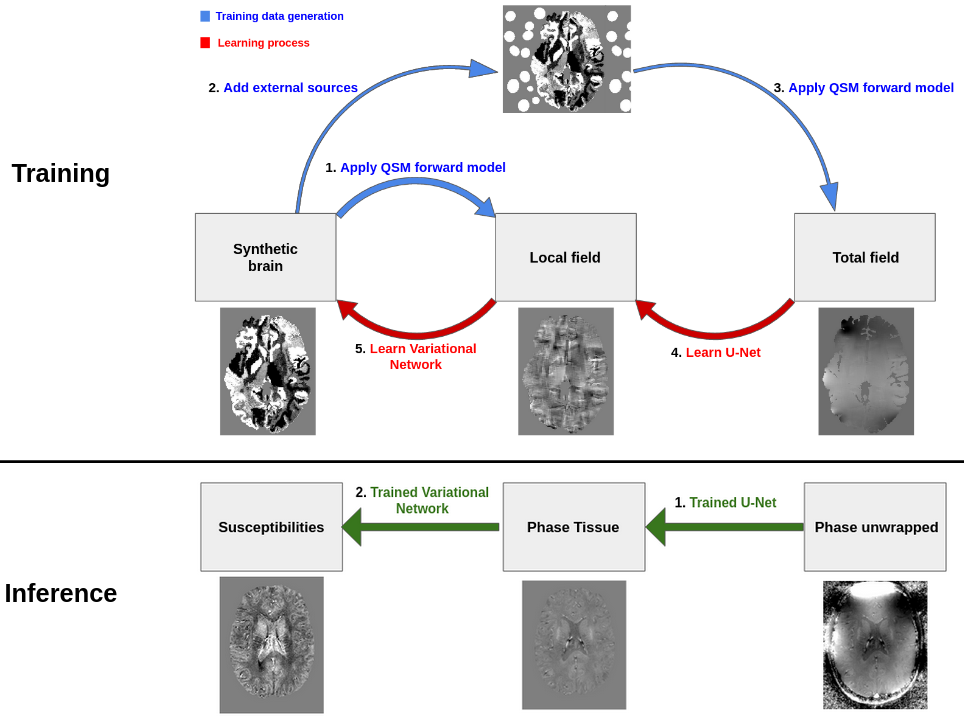}
    \caption{Illustration of the NeXtQSM pipeline for training (top) and inference time (bottom). The training includes both the training data generation process (blue) and the two deep learning models trained jointly in one optimization (red). In the training data generation (blue), we apply the QSM forward operation to the segmented brains with and without external sources to yield the data for the two learning steps. During training (red), the two architectures are trained in an end-to-end fashion. At inference time (green), the trained models perform the prediction from the unwrapped phase data to the magnetic susceptibility maps.}
    \label{fig:pipeline}
\end{figure*}

\section{Material and methods}
Our pipeline (Fig. \ref{fig:pipeline}) learns from realistic simulations of the physical properties of the problem using complex realistic structures which are synthetically created from segmentation maps, see \ref{section_training_data}.
The framework is composed of two deep learning architectures trained together in an end-to-end fashion for solving background field removal and dipole inversion. The model solving the background field removal is a standard U-Net \cite{DBLP:journals/corr/RonnebergerFB15}, whereas the dipole inversion is achieved by a Variational Network \cite{10.1007/978-3-319-66709-6_23,lai2020learned}.

\subsection{Training data}\label{section_training_data}
Due to the lack of ground truth susceptibilities for the dipole inversion and the fact that deep learning techniques need to be trained on fairly large amounts of data to be able to generalize, both of our two models were entirely trained with simulated hybrid data. \\

\begin{figure}[!h]
    \centering
    \includegraphics[width=0.7\textwidth]{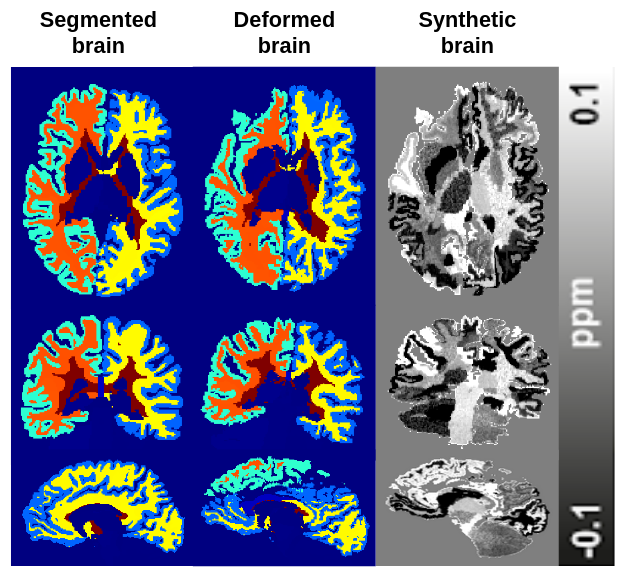}
    \caption{Visual illustration of the hybrid training data generation. The left column shows the input segmentation map, the center the deformed brain and on the right the results of the random sampling by class using a Gaussian Mixture Model.}
    \label{fig:deformedbrain}
\end{figure}

The hybrid data are based on the deformation of segmented brain MRI scans using affine transformations and a Gaussian Mixture Model \cite{billot2020learning}, which substituted the class labels with randomly sampled intensities from a prior distribution (Fig. \ref{fig:deformedbrain}). The hybrid data were generated from 28 participants (21 to 34 years of age, average of 26.5 years, 14 males) acquired on a 3 T whole-body MRI scanner (Siemens Healthcare, Erlangen, Germany). The MP2RAGE \cite{MARQUES20101271} scans (matrix = 240x256x176, resolution = 1 mm isotropic, GRAPPA = 3, TR = 4000 ms, TE = 2.89 ms, TI1/TI2 = 700/2220 ms) were segmented in 184 classes using FreeSurfer v6 \cite{FISCHL2002341}. For each of the 28 healthy individuals, 100 deformed 1 mm isotropic hybrid brains were created with different simulated susceptibility values per region drawing from the initially described normal distribution and used for training our algorithm. We tested larger amounts of training data per individual, but did not see any further training improvements. After randomly sampling the intensities per segmentation class, each brain sample was scaled to have 0 mean with a normal distribution whose 1st quartile was set to -0.5 and the 3rd to 0.5. \\

\begin{figure}[!h]
    \centering
    \includegraphics[width=0.7\textwidth]{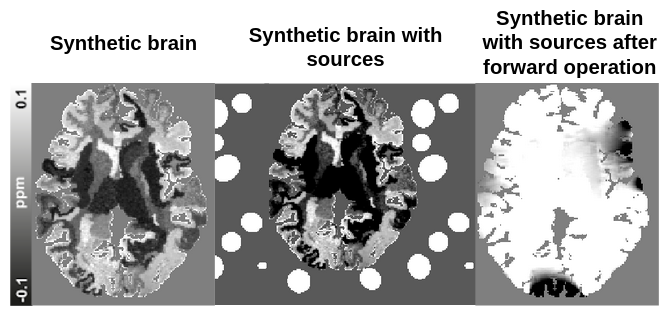}
    \caption{Illustration of the data generation for the background field removal part. Elliptical sources were randomly placed around the brain, and, after applying the QSM forward model, the background was masked out, leaving only the brain with the effect of the external sources.}
    \label{fig:backgroundfielddata}
\end{figure}

\begin{figure}[!h]
    \centering
    \includegraphics[width=0.7\textwidth]{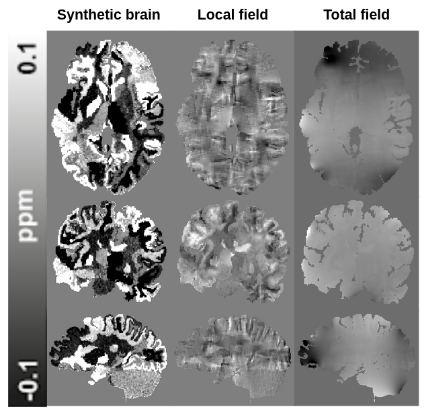}
    \caption{Illustration of the training dataset. The left column shows the initial hybrid brain with simulated susceptibility values, the center shows the data after application of the QSM dipole model and on the right the data after applying the forward model including the effect of external sources.}
    \label{fig:trainingdataset}
\end{figure}

As shown in the pipeline (Fig. \ref{fig:pipeline}), the training data generation was split into two parts: the dipole inversion and the background field removal (local field and total field respectively in Fig. \ref{fig:trainingdataset}). The dipole inversion training data were generated applying the QSM forward operation to the hybrid brains. For the background field removal problem, a random number of around 100 elliptical external sources (Fig. \ref{fig:backgroundfielddata}), simulating large susceptibility sources such as air (susceptibility value around 9.2 ppm), whose size is randomly sampled around one-tenth of the brain volume. The simulated background field sources were placed randomly around the synthetic brain before convolving it with the QSM forward model (Fig. \ref{fig:pipeline} and \ref{fig:backgroundfielddata}). \\

In our previous work (DeepQSM \cite{DBLP:journals/neuroimage/BollmannRKBOPOL19}) we showed that the QSM dipole inversion can be solved with neural networks trained purely on simulated data containing simple shapes such as cubes and spheres. However, it was shown later \cite{https://doi.org/10.1002/nbm.4292} that the simulated data in DeepQSM were not representing brain data well enough and here we improved the training data generation by using more sophisticated and complex brain-like structures that better match the real data distributions in brain scans. 

\subsection{Testing data}
It is common practice in the QSM community to compare the performance of newly developed algorithms to COSMOS \cite{COSMOS,COSMOS2} because of the high fidelity of its reconstruction. However, due to the  multiple orientations, it is actually not a good comparison to an algorithm based on a single orientation phase measurement. \\

\begin{figure}[!h]
    \centering
    \includegraphics[width=0.7\textwidth]{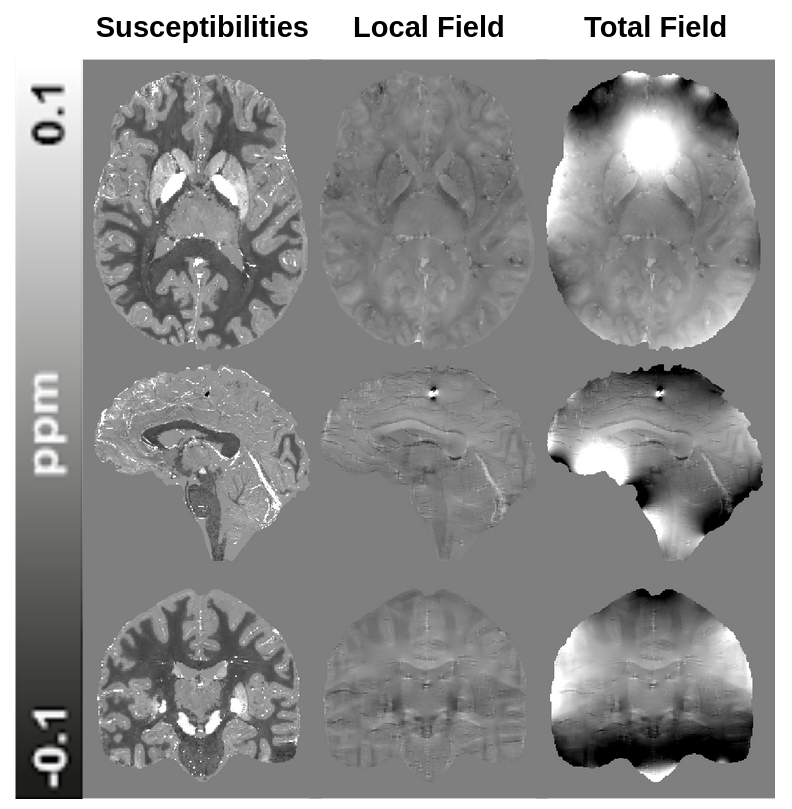}
    \caption{Sample from the dataset generated from the QSM challenge 2.0 used to test the performance of our algorithm.}
    \label{fig:testingdata}
\end{figure}

Therefore, in order to assess the quality of our reconstruction, we generated a simulated ground-truth QSM dataset using the MATLAB resources provided by the QSM challenge 2.0 \cite{https://doi.org/10.1002/mrm.28716} organizers which allow to realistically simulate the total field, local field and susceptibility. We were able to generate ground truth data for both our background field removal and dipole inversion algorithms using the SIM2 dataset of the challenge. Fig. \ref{fig:testingdata} shows a sample of the simulated testing data. For the QSM challenge testing data, the two parts of the pipeline have been tested separately using the simulated data as input. \\

Moreover, we decided to use the same tools from the QSM challenge 2.0 to generate both total and local fields at different resolutions and with different dipole kernel orientations to test the robustness of the algorithm to different input data. In Fig. \ref{fig:diffres}, the inputs at different resolutions and kernel orientations are shown as well as the predictions and corresponding ground truth. We tested our algorithm with data at 0.95 mm and 0.64 mm resolutions using both Z and Y directions of the dipole kernel. Although the training data has a fixed spatial resolution, the dipole kernel in the data term of the variational network is created accordingly to match the input data resolution, allowing the pipeline to be tested at multiple resolutions and kernel orientations. \\

Besides the data from the QSM challenge 2.0, we tested our pipeline on the second echo of an in vivo 7T scan of a 27 year old male participant. The 7T data were acquired using a multiple echo time gradient-recalled echo (GRE) 3D whole-brain data set: repetition time (TR) = 25 ms, echo time (TE) = 4.4, 7.25, 10.2, 13.25, 16.4, 19.65, 23 ms, flip angle = 13\textdegree, field of view (FOV) = $210x181.5x120 mm^3$, matrix = $280x242x160$ (0.75 mm isotropic voxels), parallel imaging (generalized autocalibrating partially parallel acquisitions (GRAPPA), acceleration factor = 2, 24 autocalibration lines), monopolar readout gradient, symmetric echo, 1116 Hz/pixel, first echo flow compensated, acquisition time (TA) = 7.9 min. For combining the individual channels we utilized the COMPOSER method \cite{https://doi.org/10.1002/mrm.26093} with the first echo of a GRE scan (3D GRE with TR = 8 ms, three echoes TE = 1.02, 3.06, 6.12 ms, flip angle = 5\textdegree, FOV = $245x245x182mm^3$, matrix = $70x70x52$ (3.5-mm isotropic voxels), monopolar readout gradient, symmetric echo, 1211 Hz/pixel, TA = 24 s). A brain mask was generated using the brain extraction tool BET from FSL \cite{JENKINSON2012782}. \\

Finally, in order to test the robustness of our pipeline, we apply Gaussian Noise perturbations with 0 mean and 0.005 variance to a control participant dataset acquired at 7T, which is one of our testing datasets used in the results section. 

\subsection{Pipeline architecture}
The first learning block (number 4 in Fig. \ref{fig:pipeline}) of the pipeline has the aim to remove background magnetic field contributions from sources which are external to the object during the MR measurement. In particular, this stage uses a U-Net \cite{DBLP:journals/corr/RonnebergerFB15} to predict the local field using a feed-forward operation given the total field. \\
The second stage uses a Variational Network \cite{lai2020learned,10.1007/978-3-319-66709-6_23} to enable a data-consistent solution of the QSM dipole inversion. The learned variational network regularizers are implemented using the U‑Net architecture \cite{DBLP:journals/corr/RonnebergerFB15} at each step. The CNN architecture adjusts its weights during training so that it helps the convergence of the reconstructed volume through a trade-off with the data term.

\subsection{Training procedure}
\begin{figure*}
    \centering
    \includegraphics[width=1\textwidth]{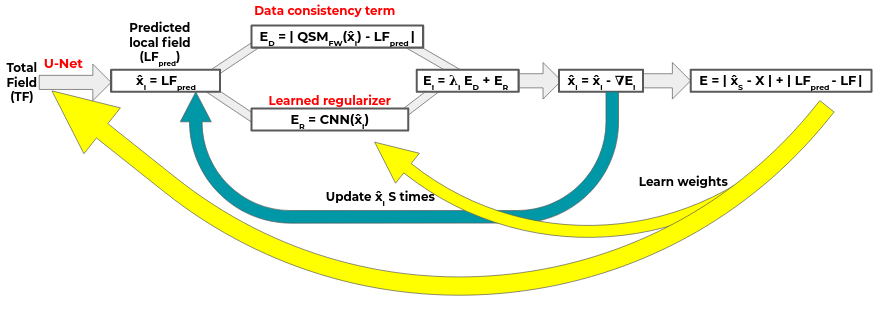}
    \caption{Graphical description of a training step. The Variational Network loss function is composed of two parts, a data consistency and a regularizer term. $X$ is the hybrid brain, $TF$ total field, $LF$ local field, $LF_{pred}$ the predicted local field, $S$ iterative steps, $\lambda$ trade-off term, $\Psi_{i}$ NeuralNet at step i, $\hat{x}_{S}$ reconstructed volume after S steps. Moreover, the backpropagation procedure is indicated by a yellow arrow which updates the weights of both the U-Net for the background field removal and the regularizer of the variational network.}
    \label{fig:modelstructure}
\end{figure*}

The training was done in an end-to-end fashion, on both the background field and dipole inversion architectures jointly after having trained them separately on our hybrid dataset. At each training epoch, 2'800 hybrid brains were used to learn the problem. The workflow of a training step can be found in Algorithm \ref{alg:trainalgorithm} and is visually illustrated in Fig. \ref{fig:modelstructure}.

Our end-to-end pipeline was implemented using Tensorflow 2.3.0 and trained for 5 days on a single NVIDIA Tesla V100 32GB GPU with batch size 2. Despite the small batch size, after initial tests we decided to not divide the volumes into smaller 3D patches and fed the complete dataset into the network to better represent the total field to the network architecture. \\
The end-to-end pipeline was trained for 100 epochs and optimized using Adam optimizer with learning rate 4e-4 and betas set to 0.9 and 0.999.

\begin{algorithm}
    \SetAlgoLined
    \DontPrintSemicolon
    $LF_{pred} \longleftarrow UNet(TF)$\;
    $\hat{x}_{i} \longleftarrow LF_{pred}$\;
    \For{$i < S$}{
        $f({\hat{x}_{i})} \longleftarrow \frac{\lambda_{i}}{N} \sum_{n=0}^{N} |LF_{pred} - \Phi \hat{x}_{i}| + \Psi(\hat{x}_{i})$\;
        $\hat{x}_{i+1} \longleftarrow \hat{x}_{i} - \frac{\partial f({\hat{x}_{i}})}{\partial \hat{x}_{i}} $\;
    }
    $E_{recon} \longleftarrow \frac{1}{N} \sum_{n=0}^{N} |\hat{x}_{S} - X| + |LF_{pred} - LF|$\;
    min $E_{recon}$
    \caption{NeXtQSM end-to-end training step. $X$ is hybrid brain, $TF$ total field, $LF$ local field, $LF_{pred}$ the predicted local field, $S$ iterative steps, $\lambda$ trade-off term, $\Psi$ NeuralNet, $N$ batch size, $\hat{x}_{S}$ reconstructed volume after S steps. The arrows indicate the assignment to a new variable.}
    \label{alg:trainalgorithm}
\end{algorithm}

\subsection{Evaluation}
Our end-to-end deep learning pipeline to solve background field removal and dipole inversion was compared both quantitatively on metrics and visually to other methods which offer the solution of these two QSM steps together. Although the pipeline is trained end-to-end, it is possible to visualize intermediate results such as the output of the background field removal before being fed into the dipole inversion. \\

Therefore, we compare background field removal algorithms (LBV \cite{https://doi.org/10.1002/nbm.3064}, V-SHARP \cite{https://doi.org/10.1002/nbm.3550}, Harperella \cite{https://doi.org/10.1002/nbm.3056} and SHARQnet \cite{BOLLMANN2019139}), dipole inversion algorithms (iLSQR \cite{li_method_2015}, Star-QSM \cite{https://doi.org/10.1002/nbm.3383}, DeepQSM \cite{DBLP:journals/neuroimage/BollmannRKBOPOL19}, FANSI \cite{https://doi.org/10.1002/mrm.27073}) as well as algorithms which solve consecutive steps jointly (TGV-QSM \cite{tgvqsm}, AutoQSM \cite{autoqsm} and a mix of different background field and dipole inversion algorithms). The final results for FANSI and TGV-QSM have been chosen after fine-tuning the regularization parameters using the L-curve procedure (alpha for FANSI \cite{https://doi.org/10.1002/mrm.27073} and TGV-QSM \cite{tgvqsm}). \\
The results have been compared to other methods using the quantitative metrics NRMSE, HFEN and XSIM \cite{xsim}.

\section{Results}
\subsection{QSM Reconstruction Challenge 2.0}
\begin{figure*}[!h]
    \centering
    \includegraphics[width=1.0\textwidth]{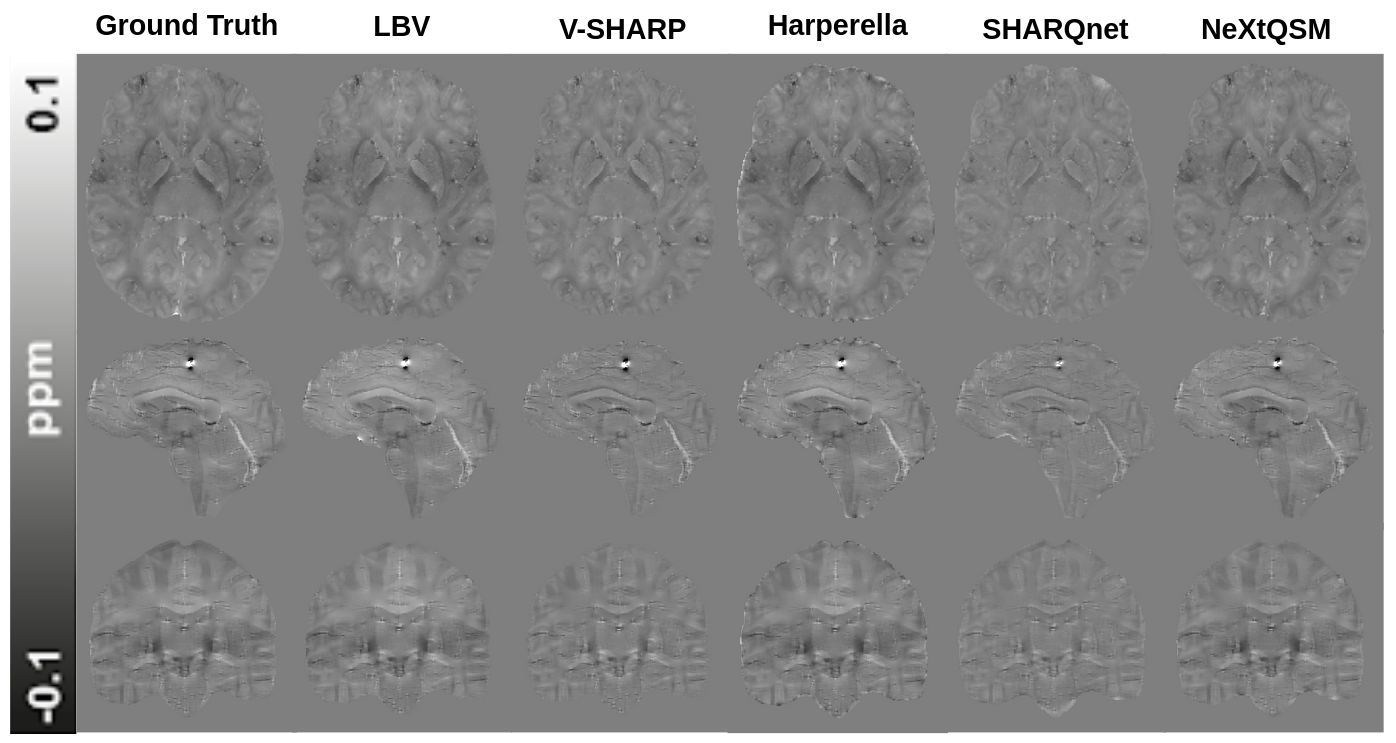}
    \caption{Visual comparison of different methods for removing the background field using the QSM challenge 2.0 data.}
    \label{fig:bfqsm}
\end{figure*}

\begin{figure*}[!h]
    \centering
    \includegraphics[width=1.0\textwidth]{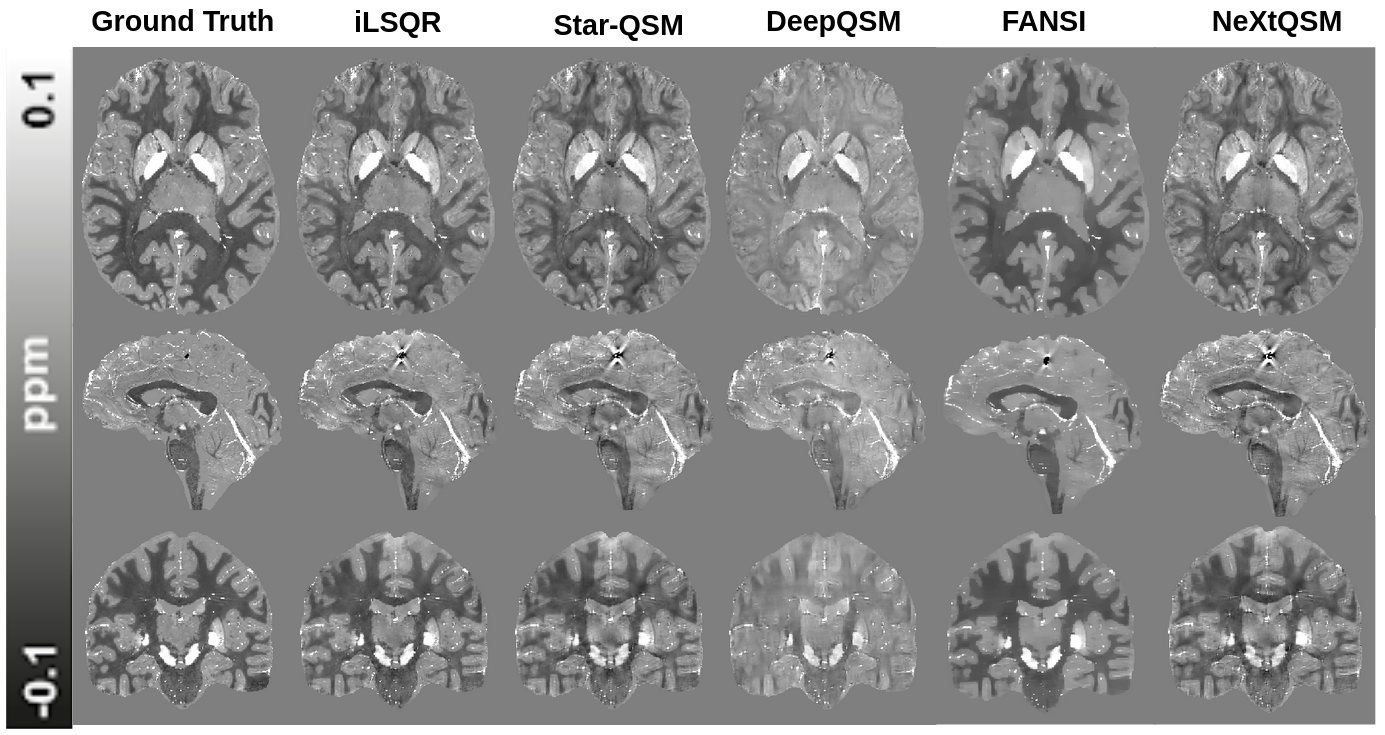}
    \caption{Visual comparison of different methods for dipole inversion using the QSM challenge 2.0 data.}
    \label{fig:dipqsm}
\end{figure*}

The first experiment to evaluate the performance of the pipeline was performed on data simulated with tools provided by the QSM challenge 2.0 \cite{https://doi.org/10.1002/mrm.28716}. \\
Fig. \ref{fig:bfqsm} shows the result of our background field removal algorithm compared to alternative methods as well as quantitative metrics computed with respect to the simulated ground truth and displayed in Table \ref{tab:bfmetric}. It can be observed both visually and quantitatively that our background field removal is able to perform on a comparable level to the other considered methods. \\
Fig. \ref{fig:dipqsm} and Table \ref{tab:dipmetric} compare our dipole inversion algorithm to other established dipole inversion algorithms. Results of our end-to-end pipeline are compared in the next paragraph with in vivo data of a healthy participant.

\begin{table}[!htb]
\centering
\caption{\label{tab:bfmetric}Quantitative comparison of background field removal methods to simulated local field from QSM challenge 2.0.}
\begin{tabular}{ |p{2.5cm}||p{1.3cm}|p{1.3cm}|p{1.3cm}|}
    \hline
    \multicolumn{4}{|c|}{\textbf{Background Field Removal}} \\
    \hline
    \textbf{Method}  & \textbf{NRMSE}    &\textbf{XSIM}   &\textbf{HFEN}\\
    \hline
    LBV (MEDI)  &51.64   &\textbf{0.86}    &\textbf{31.20}\\
    \hline
    V-SHARP  &61.58  &0.78    &37.05\\
    \hline
    Harperella  &60.74  &0.83   &68.74\\
    \hline
    SHARQnet  &68.93  &0.74   &45.43\\
    \hline
    NeXtQSM  &\textbf{50.21}  &0.83   &36.87\\
    \hline
\end{tabular}
\end{table}

\begin{table}[!htb]
\centering
\caption{\label{tab:dipmetric}Quantitative comparison of dipole inversion reconstruction methods from the simulated local field of the QSM challenge 2.0.}
\begin{tabular}{ |p{2.5cm}||p{1.3cm}|p{1.3cm}|p{1.3cm}|}
    \hline
    \multicolumn{4}{|c|}{\textbf{Dipole Inversion}} \\
    \hline
    \textbf{Method}  & \textbf{NRMSE}    &\textbf{XSIM}    &\textbf{HFEN}\\
    \hline
    iLSQRT  &45.31   &\textbf{0.76}     &49.94\\
    \hline
    Star-QSM  &52.15  &0.65   &59.22\\
    \hline
    DeepQSM  &84.33  &0.30 &69.26\\
    \hline
    FANSI  &\textbf{41.83}  &0.66 &\textbf{43.06}\\
    \hline
    NeXtQSM  &60.81  &0.65 &60.81\\
    \hline
\end{tabular}
\end{table}

\begin{figure*}[!h]
    \centering
    \includegraphics[width=0.9\textwidth]{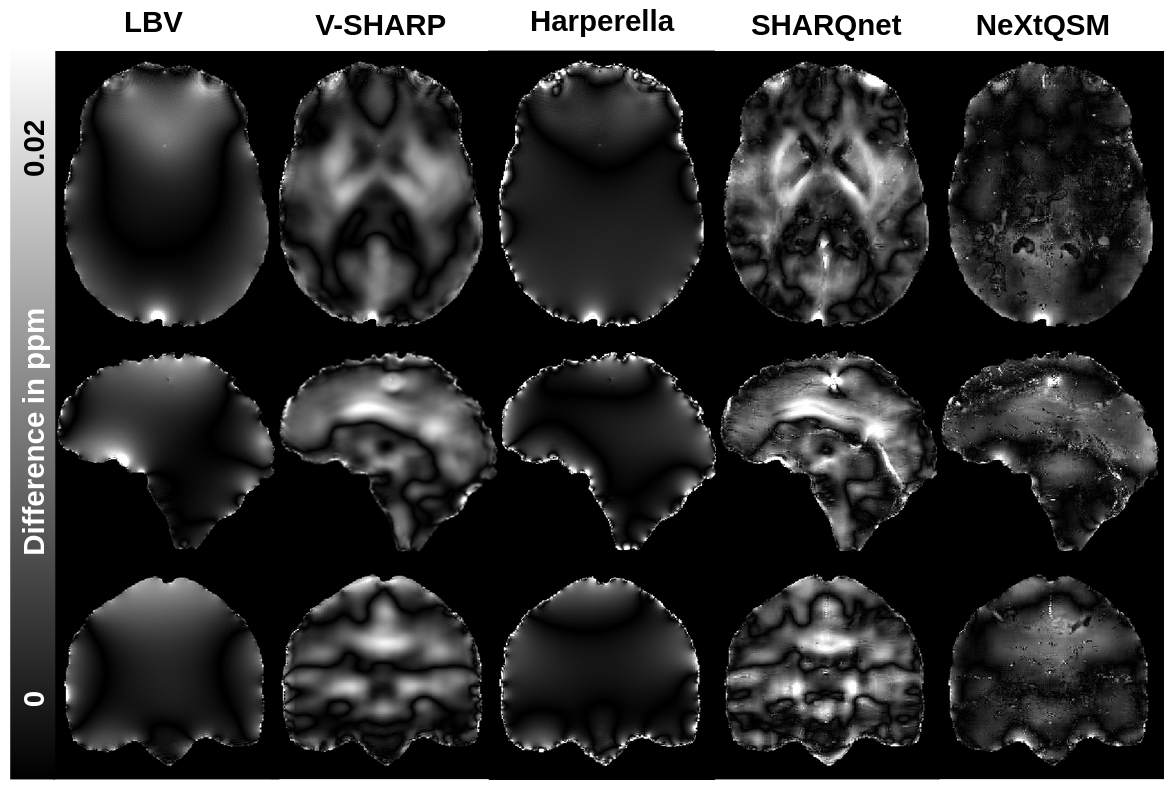}
    \caption{Difference image of background field removal methods with the simulated ground truth.}
    \label{fig:diffebfqsm}
\end{figure*}

\begin{figure*}[!h]
    \centering
    \includegraphics[width=0.9\textwidth]{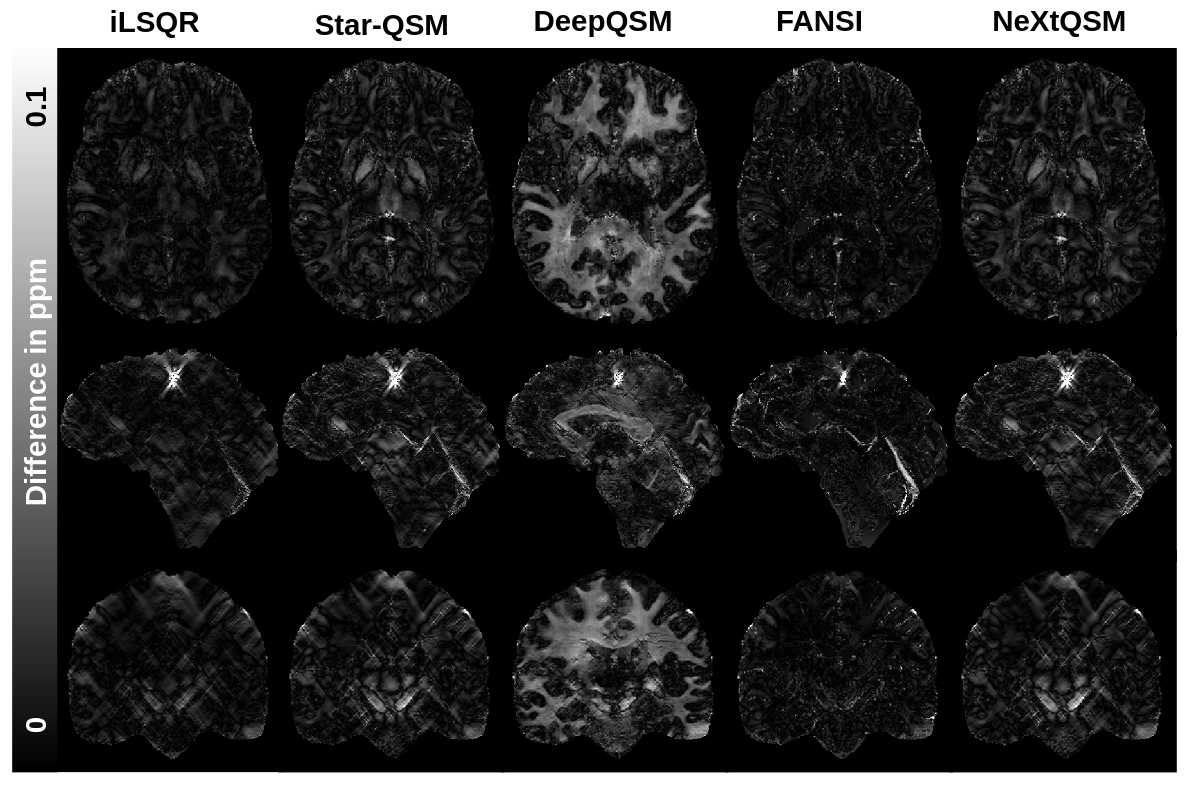}
    \caption{Difference image of dipole inversion methods with the simulated ground truth.}
    \label{fig:diffedipqsm}
\end{figure*}

\subsection{Different resolutions and dipole kernel orientations}
\begin{figure*}[!h]
    \centering
    \includegraphics[width=1.0\textwidth]{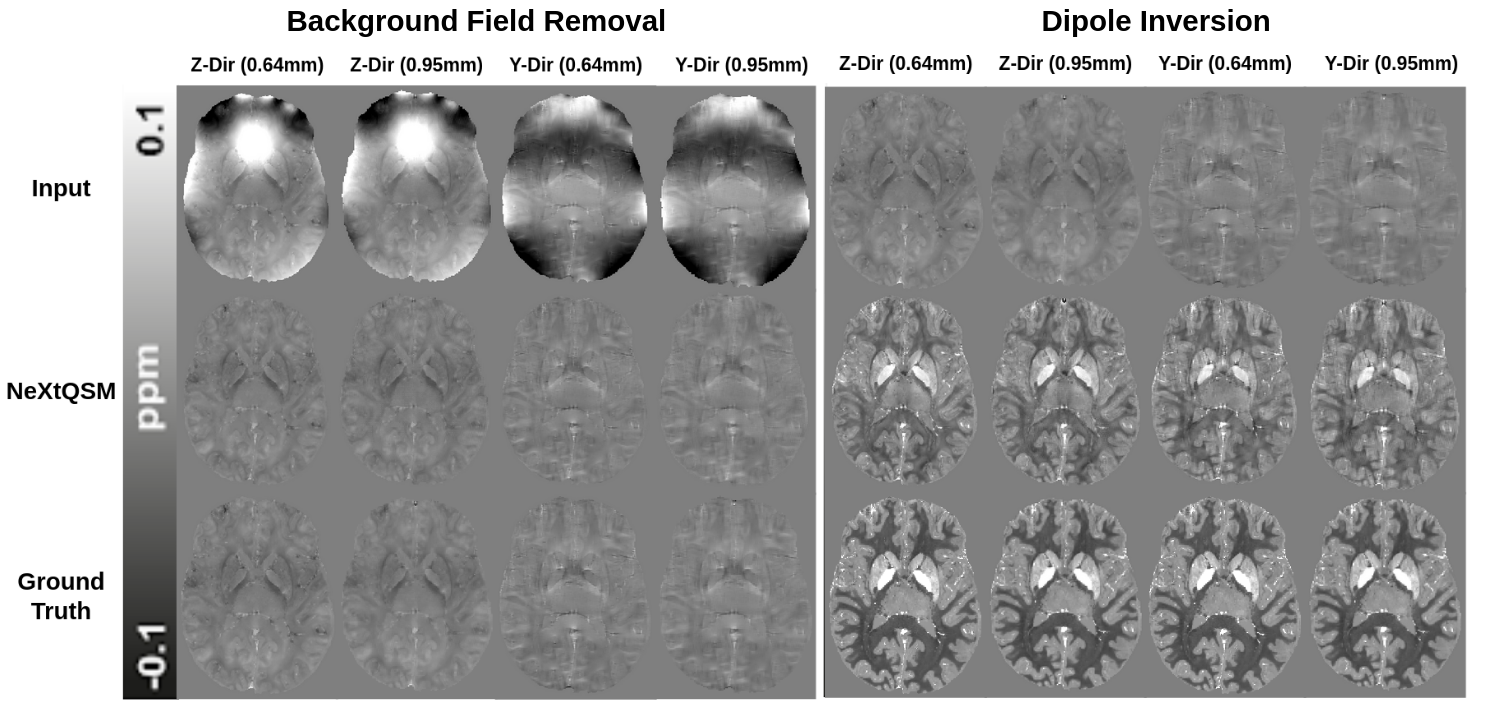}
    \caption{Visual comparison of NeXtQSM on different resolutions and directions of the dipole kernel. In the first row, the input of the corresponding part is shown, in the second row the output predicted by our model, whereas the last row indicates the ground truth that should have been reconstructed.}
    \label{fig:diffres}
\end{figure*}

The results shown in Fig. \ref{fig:diffres} demonstrate that NeXtQSM is robust to different resolutions and dipole kernel orientations.

\subsection{Healthy subject dataset}
\begin{figure*}[!h]
    \centering
    \includegraphics[width=1.0\textwidth]{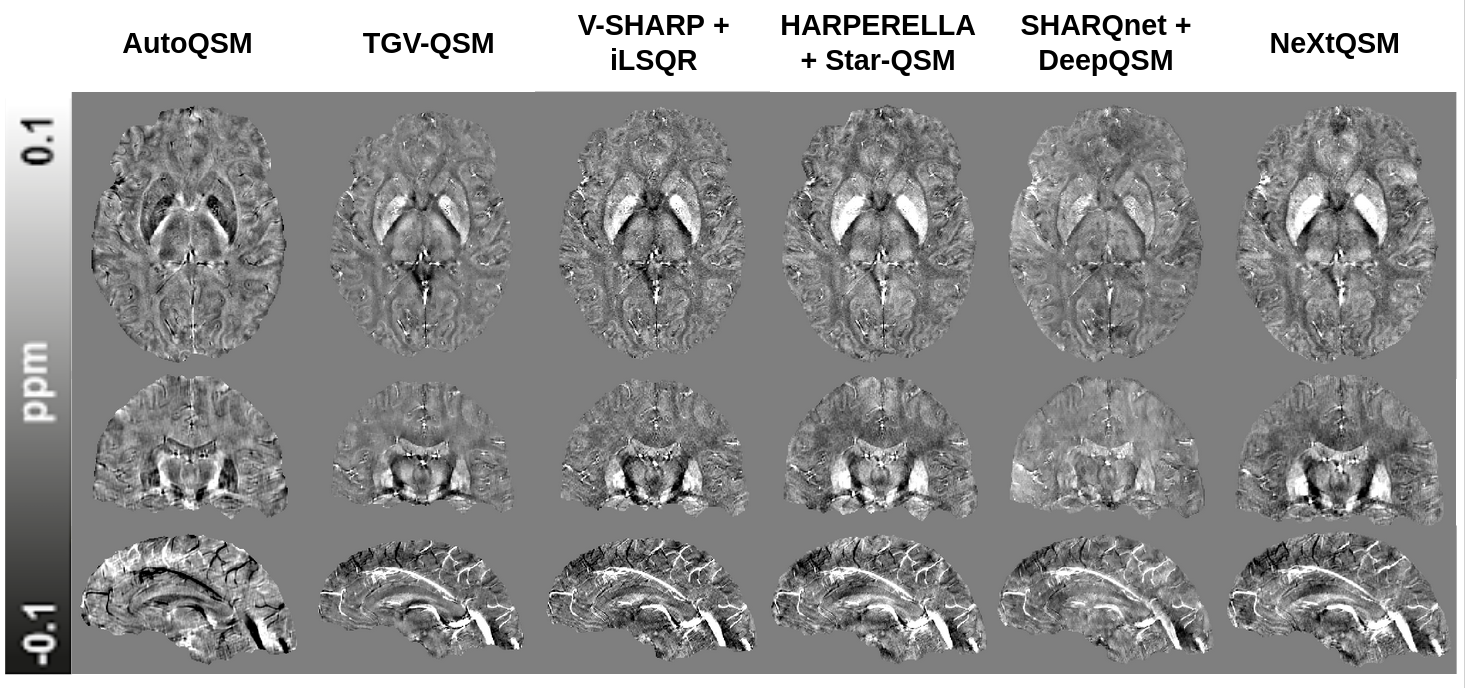}
    \caption{Visual comparison of different methods using the 7 T control participant dataset. It can be observed that all methods deliver artifact free QSM reconstructions, but SHARQnet and DeepQSM show reduced contrast in comparison to the other methods.}
    \label{fig:resultshealthy}
\end{figure*}

A second experiment utilized a control participant dataset which is shown in Fig. \ref{fig:resultshealthy} to show that our end-to-end pipeline is applicable to ultra-high field MRI data. The visual comparison indicates that NeXtQSM delivers a robust background field correction and dipole inversion similar to established QSM methods like TGV-QSM, STI Suite and AutoQSM. It can also be observed that the model agnostic deep learning QSM method SHARQnet + DeepQSM  suffers from a lack in contrast and shows residual background field artifacts.

\subsubsection{Robustness to Gaussian noise perturbations}
\begin{figure*}[!h]
    \centering
    \includegraphics[width=1.0\textwidth]{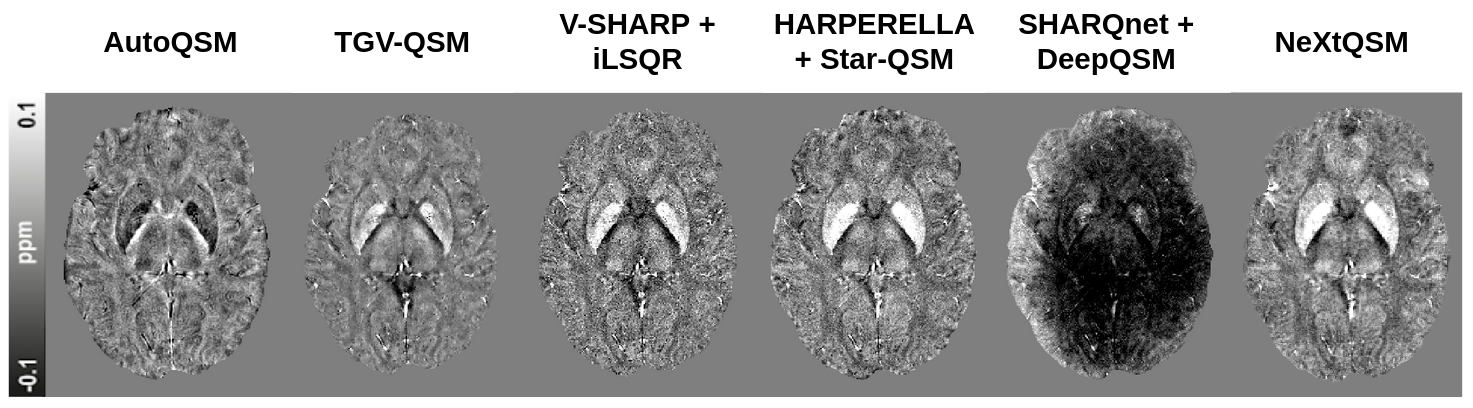}
    \caption{Visual comparison of the robustness to Gaussian noise perturbations applied to the input. It can be seen that model-agnostic deep learning like SHARQnet and DeepQSM fail to be robust to a change in input data distribution.}
    \label{fig:noisecomparison}
\end{figure*}

It can be observed in Fig. \ref{fig:noisecomparison} that the data consistency constraint in NeXtQSM has a stabilising impact against Gaussian noise perturbation especially in comparison to other deep learning approaches such as SHARQnet + DeepQSM. \\

We also measured the processing times of the different pipelines (tested on an Intel Core i5-10300H CPU) and the run-times range from  55 s for AutoQSM \cite{autoqsm}, 500 s for TGV-QSM \cite{tgvqsm}, 160 s for V-Sharp \cite{https://doi.org/10.1002/nbm.3550} and iLSQR \cite{li_method_2015}, 80 s for Harperella \cite{https://doi.org/10.1002/nbm.3056} and Star-QSM \cite{https://doi.org/10.1002/nbm.3383}, 70 s for SHARQnet \cite{BOLLMANN2019139} and DeepQSM \cite{DBLP:journals/neuroimage/BollmannRKBOPOL19} and 90s for NeXtQSM.

\section{Discussion}
The proposed NeXtQSM pipeline is capable of learning the background field correction and dipole inversion from hybrid training data. We show that the trained networks generalize well to simulated data from the QSM challenge 2.0 and a 7T ultra-high field MRI dataset. When comparing to established QSM algorithms both learning and non-learning based, NeXtQSM's hybrid training data was shown to capture both the geometrical and physical properties needed to train deep learning models. The data enabled NeXtQSM to learn the QSM dipole model and the effect of external non-brain sources causing the background field. NeXtQSM's integrated training approach resulted in a pipeline that delivered fast and robust results. Inclusion of the data consistency constraint helped increasing the robustness of the processing to noise and enabled NeXtQSM to generalize well to in vivo data.\\

A key difference to existing deep learning QSM solutions is that our framework includes a data consistency constraint for the dipole inversion implemented via a variational network. This increased robustness can be seen for example via a perturbation with Gaussian Noise of the input data resulting in unusable results for a model agnostic deep learning methods, such as DeepQSM and SHARQnet (see Fig. \ref{fig:noisecomparison}). In addition, our pipeline is also robust to a change in resolution and directions of the dipole kernel. \\

Contrary to earlier work \cite{chen2019qsmgan,DBLP:journals/neuroimage/BollmannRKBOPOL19,BOLLMANN2019139} training NeXtQSM with larger patches was beneficial and did not need to be limited to a training regime that would split-up the brain-volumes in smaller sub-patches. With state-of-the-art GPU hardware we managed to train NeXtQSM with the complete brain volume using a small batch-size. This training regime better matches the theoretical concepts underlying QSM. The background field correction especially profits from a "complete picture" of the artifacts caused by external sources, like air-tissue interfaces. \\

Although other deep learning approaches have previously attempted to solve the background field removal and dipole inversion in one step \cite{autoqsm,liu2019deep}, these methods were very sensitive to slight perturbations in the input data. The training data used were generated from established methods from acquired in vivo datasets which may have biased the learning. The actual physical model utilized to construct our hybrid data means that we could create a large, controllable and independent data set to train on. \\

Another advantage of NeXtQSM is that it is able to run without any fine-tuning of parameters like in traditional iterative methods. The need of a regularization parameter fine-tuning during inference is a non-negligible limiting factor for QSM clinical applications and clinicians who want a tool which is automated and robust. \\

Deep learning techniques deliver very fast and high fidelity QSM solutions but the lack of data consistency constraints makes them very sensitive to data distribution changes. Other approaches like TGV-QSM \cite{tgvqsm} deliver data consistent solutions, but the optimization process is computationally expensive (500s for TGV QSM and 160s for V-Sharp and iLSQR in STI-Suite) limiting their potential clinical applications. NeXtQSM elegantly combines the advantages of deep learning with data consistency concepts used in established QSM inversion techniques to be substantially faster (90 s) than traditional techniques but more robust to input data variations than other  deep learning approaches. \\

Although NeXtQSM delivers quantitative results that are in line with traditional dipole inversion methods, it is not outperforming traditional methods in accuracy or image contrast and we opted for a final result that does not produce over-regularized maps. Another limitation of NeXtQSM is that it does not suppress the calcification streaking artifact, which is in line with the results of traditional methods like iLSQR and Star-QSM. 

\section{Conclusion}
In this study, we presented NeXtQSM, a deep learning pipeline trained on hybrid samples that integrates and solves QSM’s background field and dipole inversion steps in a data consistent fashion. Our integrated approach is more robust to variations in the input data than other deep learning methods, and substantially faster than iterative QSM techniques.

\section*{Acknowledgments}
The authors acknowledge the facilities and scientific and technical assistance of the National Imaging Facility, a National Collaborative Research Infrastructure Strategy (NCRIS) capability, at the Centre for Advanced Imaging, the University of Queensland. This research was undertaken with the assistance of resources and services from the Queensland Cyber Infrastructure Foundation (QCIF). Oracle for Research provided Oracle Cloud credits and related resources to support this project.  MB acknowledges funding from Australian Research Council Future Fellowship grant FT140100865. SR was supported by the Marie Skłodowska-Curie Action (MS-fMRI-QSM 794298). This research was funded (partially or fully) by the Australian Government through the Australian Research Council (project number IC170100035). Funding for the position of F.B.L. by the German Research Foundation is gratefully acknowledged (LA 2804/12-1).

\section*{Conflict of Interest Statement}
The presented method has been submitted as a patent application and two co-authors (Kieran O’Brien and Jin Jin) are employees of Siemens Healthcare Pty Ltd.

\bibliographystyle{plain}
\bibliography{samplepaper.bib}

\end{document}